\documentclass[a4paper, 11pt]{article}
\usepackage{amsmath,amsfonts,amssymb}
\usepackage{amssymb,epsf}
\usepackage{graphicx}
\usepackage{latexsym}

\begin{document}

\title{Nielsen-Olesen vortices for large Ginzburg-Landau parameter}
\author{{\large J{\" u}rgen Burzlaff}$^{\diamond \star}$
and {\large F. Navarro-L{\' e}rida}$^{\dagger}$
\\
\\
$^{\diamond}${\small School of Mathematical Sciences, Dublin City
University, Ireland}
\\
$^{\dagger}${\small Departamento de F{\' i}sica At{\' o}mica, Molecular y Nuclear, Ciencias F{\' i}sicas,}
\\{\small Universidad Complutense de Madrid, Spain}
\\
$^{\star}${\small School of Theoretical Physics, Dublin Institute for Advanced Studies, Ireland }}

\maketitle

\begin{abstract}
Using analytic and numerical techniques Nielsen-Olesen vortices, which in the context of Ginzburg-Landau theory are known as Abrikosov vortices of type-II superconductors, are studied for large Ginzburg-Landau parameter $\lambda$. We show that their energy is equal to $(\pi n^2 /2)\log\lambda$ to leading order, where $n$ is the winding number of the vortex, and find that the limit of the gauge field can be expressed in terms of the modified Bessel function $K_1$. The leading terms of the asymptotic expansion of the solution are given, and the different contributions to the energy are analyzed. \\

\end{abstract}

%%%%%%%%%%%%%%%%%%%%%%%%%%%%%%%%%%%%%%%%%%%%%%%%%%%%%%%%%%%%%%%%
\section{Introduction}
%%%%%%%%%%%%%%%%%%%%%%%%%%%%%%%%%%%%%%%%%%%%%%%%%%%%%%%%%%%%%%%%

Of all the localized finite-energy solutions of classical gauge theories, the vortices of the Abelian Higgs model in 2 space dimensions, the prototype of a gauge field theory with spontaneous symmetry breaking, should be the ones easiest to understand. However, none of these solutions is given in terms of known functions. Nielsen and Olesen~\cite{Nielsen-Olesen} found the time-independent, radially symmetric, localized finite-energy solutions of the Abelian Higgs model in 2 space dimensions, the Nielsen-Olesen vortices, by reducing the equations of motion to two second-order equations for two radial functions. The mathematically rigorous proof that the resulting equations for the two radial functions have solutions with the required properties was given by Tyupkin et al.~\cite{Tyupkin},  and Berger and Chen~\cite{Berger-Chen}.

In the context of Ginzburg-Landau theory, which is the time-independent Abelian Higgs model without an electric field, the Nielsen-Olesen vortices are known as Abrikosov vortices of type-II superconductors~\cite{Abrikosov}. This means that the properties of Nielsen-Olesen vortices can, and have been, studied in experiments. The Nielsen-Olesen vortices also provide a simple example of cosmic strings~\cite{Hindmarsh-Kibble}, which might explain some of the structures seen in the universe today.

With the solution not available in terms of known functions, numerical
computations become all the more important. For the Nielsen-Olesen vortices
the numerical work started soon after the solutions were
found~\cite{Jacobs-Rebbi}~\cite{Hill}. Asymptotic analysis is another technique often applied when the explicit solution is not known. For the Nielsen-Olesen vortices, Berger and Chen~\cite{Berger-Chen} obtained some asymptotic results for large Ginzburg-Landau parameter.  The asymptotic analysis of the monopole structure was given by Kirkman and Zachos~\cite{Kirkman-Zachos}. More recently, the same techniques were used for the Skyrmion~\cite{Brihaye 1} and a Skyrme-like monopole~\cite{Brihaye 2}. In this paper, we perform a similar asymptotic analysis for the Nielsen-Olesen vortices.

%%%%%%%%%%%%%%%%%%%%%%%%%%%%%%%%%%%%%%%%%%%%%%%%%%%%%%%%%%%%%%%%%%%%%%%%%%%%%%
\section{Radially symmetric solutions}
%%%%%%%%%%%%%%%%%%%%%%%%%%%%%%%%%%%%%%%%%%%%%%%%%%%%%%%%%%%%%%%%%%%%%%%%%%%%%%

The Hamiltonian density of the time-independent Abelian Higgs model in 2 space dimensions is given by
\begin{equation}\label{EQ-H}
{\cal H} = \frac{1}{4}F_{ij} F^{ij} + \frac{1}{2}(D_i \phi)(D^i
\phi)^* + \frac{\lambda}{8}(\mid \phi \mid ^{2} -1)^{2} .
\end{equation}
Here $D_i \phi =\partial_i \phi - \imath A_i \phi$ and $F_{ij} = \partial_i A_j -
\partial_j A_i$ ($i$,$j$=1,2) are the covariant derivative and the field strength, respectively, and the metric is $g={\rm diag}(1,1)$. ${\cal H}$ in Eq.~(\ref{EQ-H}) is also the Ginzburg-Landau free energy of a superconductor. In this model, the Ginzburg-Landau parameter $\lambda$ is equal to 1 at the point between type-I and type-II superconductivity. The corresponding Euler-Lagrange equations are
\[
D_i D^i \phi - \frac{\lambda}{2} \phi (|\phi |^2 -1 ) = 0 ,
\]
\begin{equation}
\partial_i F^{ji} + \frac{\imath}{2}
\left[ \phi ^* D^j \phi -  \phi
(D^j \phi )^*\right] = 0 .
\end{equation}

The Euler-Lagrange equations have radially symmetric solutions of the form
\begin{equation}
\phi = f (r) e^{\imath n\theta}, \quad A_i = - \frac{
a(r)}{r^2}\varepsilon_{ij} x^j ,
\end{equation}
where $n=\pm 1,\pm 2, ...$ is the winding number.
The radial functions satisfy the equations
\begin{equation}\label{EQ-af}
a'' - \frac{1}{r} a' +f^2 (n-a) = 0 ,
\quad f'' + \frac{1}{r} f' -\frac{(n-a)^2}{r^2} f = \frac{\lambda}{2} (f^2 - 1) f ,
\end{equation}
and the boundary conditions for regular vortex solutions to exist are
\begin{equation} \label{bcs}
f(0)=a(0)=0, \quad \lim_{r \rightarrow \infty}f(r)=1, \quad \lim_{r \rightarrow
\infty} a(r) = n .
\end{equation}
These solutions are the Nielsen-Olesen vortices~\cite{Nielsen-Olesen} of the Abelian Higgs model; or for $\lambda >1$ the Abrikosov vortices of type-II superconductors.
The existence proof for such solutions was given by Tyupkin et al.~\cite{Tyupkin}.
The proof is based on the fact that the Nielsen-Olesen solution minimizes the energy
\begin{equation}\label{EQ-E}
E[a(r),f(r)] = \int_0^\infty {\cal E} \;dr = 2\pi \int_0^\infty \left[
  \frac{a'^2}{2r} +\frac{r}{2}f'^2 + \frac{1}{2r} (n-a)^2 f^2 +
  \frac{\lambda}{8} r (f^2 -1)^2 \right] \; dr .
\end{equation}
The asymptotic behaviour of the solutions for $r\ll 1$ (and finite $\lambda$) is
\begin{equation}\label{EQ-Asym1}
f=f_n r^n - \frac{(\lambda + 4na_2 )f_n}{8(n+1)} r^{n+2} + \dots , \quad
a=a_2 r^2 - \frac{f_n^2}{4(n+1)} r^{2n+2} + \dots \ . 
\end{equation}
For $r\gg 1$ we have~\cite{Perivolaropoulos}
\begin{equation}\label{EQ-Asym2}
a= n + \alpha \sqrt{r} {\rm e}^{-r} + \dots , \quad
f= \left\{ \begin{array}{cl}
1 + \beta \frac{{\rm e}^{-\sqrt{\lambda}r}}{\sqrt{r}} + \dots
& (\lambda\leq 4)\\
1 + \frac{\alpha^2 {\rm e}^{-2r}}{(4-\lambda )r} + \dots
& (\lambda > 4) \end{array}\right. .
\end{equation}

Equations (\ref{EQ-af}) with boundary conditions Eq.~(\ref{bcs}) cannot be solved analytically. By employing a collocation method for boundary-value ordinary
differential equations equipped with an adaptive mesh selection procedure in a
compactified grid
\cite{colsys},  we have solved numerically the equations with high accuracy (global
tolerance  $10^{-9}$) for a large range of values of $\lambda$. In Fig.~1 we
show the energy $E$ as a function of $\lambda$ for small values of
$\lambda$. We clearly see that $E/n$ does not depend on $n$ at $\lambda=1$, and is
increasing with $n$ for $\lambda>1$ and
decreasing with $n$ for $\lambda<1$. That $E/n$
does not depend on $n$ means that the forces balance at $\lambda=1$, which makes it possible for solutions corresponding to vortices at arbitrary separation to exist~\cite{Taubes}.

\begin{figure}[h!]
\epsfxsize=7cm
\centerline{\includegraphics[angle=0,width=92mm]{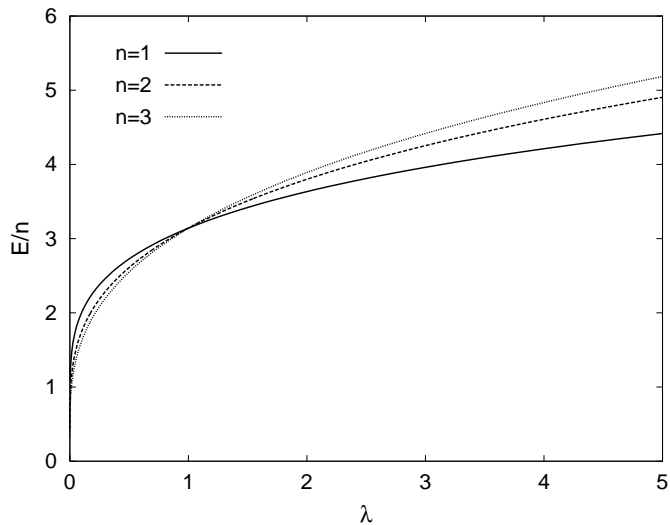}}
\caption{
Energy per vortex number $E/n$ versus $\lambda$ for Nielsen-Olesen solutions
with $n=1,2,3$.
}
\label{fig1}
\end{figure}

Extending the computations for larger values of $\lambda$ we observe numerically
a logarithmic divergence of the energy. This is exhibited in Fig.~2. One can
also see that, to leading order, the energy increases quadratically with the vortex number $n$.
A detailed analysis of the numerical data reveals that the energy follows the
following asymptotic formula
\begin{equation}
\frac{E^{\rm num}}{n^2} = \frac{\pi}{2} \log\lambda + \Delta(n) + o(1) \ , \label{eq_E_num}
\end{equation}

\begin{figure}[h!]
\epsfxsize=7cm
\centerline{\includegraphics[angle=0,width=92mm]{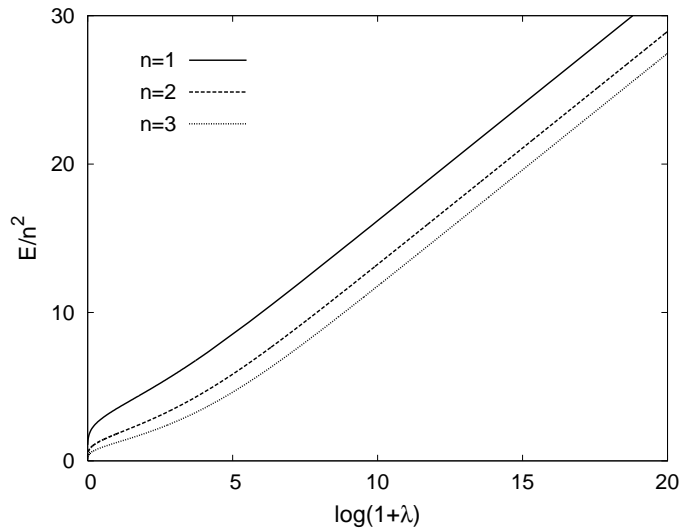}}
\caption{
Energy over $n^2$ versus $\log(1+\lambda)$ for Nielsen-Olesen solutions
with $n=1,2,3$.
}
\label{fig2}
\end{figure}

\noindent where
\begin{equation}
\Delta(n) = \left\{
\begin{array}{rl}
0.47199 , & n=1 \\
-2.48172 , & n=2 \\
-3.95372 , & n=3
\end{array}
\right. \ . \label{Delta}
\end{equation}
In the next section we will prove rigorously that the energy behaves like that by
performing an asymptotic analysis of Nielsen-Olesen solutions.

%%%%%%%%%%%%%%%%%%%%%%%%%%%%%%%%%%%%%%%%%%%%%%%%%%%%%%%%%%%%%%%%%%%%%%%%%%%%%%
\section{Asymptotic analysis}
%%%%%%%%%%%%%%%%%%%%%%%%%%%%%%%%%%%%%%%%%%%%%%%%%%%%%%%%%%%%%%%%%%%%%%%%%%%%%%

For our asymptotic analysis, we first split the energy Eq.~(\ref{EQ-E}) into four parts,

\begin{eqnarray}
&&E_1 = 2\pi \int_0^\infty \frac{a'^2}{2r} \; dr \ , \\
&&E_2 = 2\pi \int_0^\infty  \frac{r}{2}f'^2 \; dr \ , \\
&&E_3 = 2\pi \int_0^\infty \frac{1}{2r} (n-a)^2 f^2 \; dr \ , \\
&&E_4 = 2\pi \int_0^\infty \frac{\lambda}{8} r (f^2 -1)^2   \; dr \ . \label{Es}
\end{eqnarray}

\noindent These four contributions to the total energy correspond to the gauge field
contribution ($E_1$), the Higgs dynamical contribution ($E_2$ and $E_3$), and
the contribution of the potential ($E_4$), respectively.

To study the dependence of the energy on $\lambda$ we differentiate
Eq.~(\ref{EQ-E}) with respect to $\lambda$ and obtain
\[ \frac{dE}{d\lambda} = \int_0^\infty \left( \frac{\partial
{\cal E}}{\partial\lambda} + \frac{\partial a}{\partial \lambda}
\frac{\partial {\cal E}}{\partial a} + \frac{\partial
a'}{\partial \lambda}\frac{\partial {\cal E}}{\partial a'} +
\frac{\partial f}{\partial \lambda}\frac{\partial {\cal E}}{\partial f}
+ \frac{\partial f'}{\partial \lambda}
\frac{\partial {\cal E}}{\partial f'} \right) dr \]
\begin{equation}\label{EQ-dE}
= \int_0^\infty \frac{\pi r}{4} (f^2 -1)^2 \; dr > 0 .
\end{equation}
Here we have used
integration by parts, the equations for $a$ and $f$ Eq.~(\ref{EQ-af}), and have assumed that $(a'/r)(\partial a /\partial\lambda )$ and $rf'(\partial f /\partial\lambda )$
vanish as $r\rightarrow 0$ and as $r\rightarrow\infty$. We see that the energy increases with
$\lambda$, and, if the energy is bounded, that $f =1$ ($r>0$) in the limit $\lambda\rightarrow\infty$.

We will now show that the energy of the Nielsen-Olesen vortex is not bounded
for $\lambda\rightarrow\infty$, but that nevertheless $f$ will approach the
singular limit $f=1$ ($r>0$). We start by considering both possibilities. If
we do not have $f=1$ ($r>0$) in the limit, the integral in Eq.~(\ref{EQ-dE}) does
not go to zero and $E_4$ is at least of order $\lambda$ for large $\lambda$.
That is in contradiction to numerics. In Fig.~3 we exhibit $E_4$ as a
function of $\lambda$. It clearly tends to $n^2 \pi/2$ in the limit $\lambda
\to \infty$, so it is bounded in that limit.

\begin{figure}[h!]
\epsfxsize=7cm
\centerline{\includegraphics[angle=0,width=92mm]{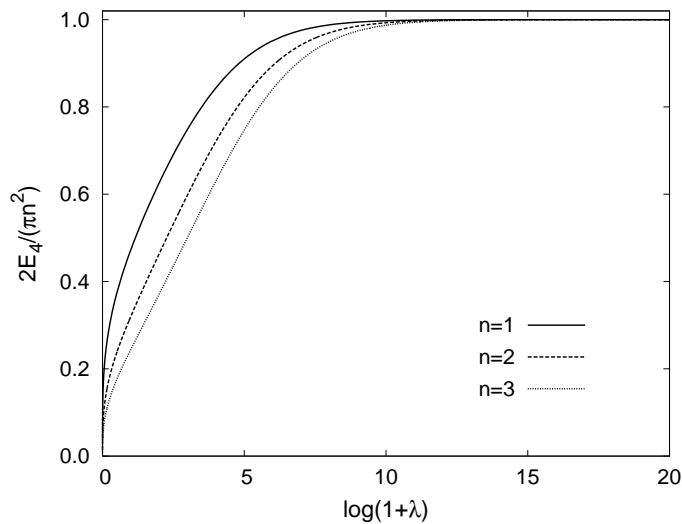}}
\caption{
$E_4$ versus $\log(1+\lambda)$ for Nielsen-Olesen solutions
with $n=1,2,3$.
}
\label{fig3}
\end{figure}

\noindent On numerical evidence, we conclude that function $f$ tends to the singular limit $f=1$ ($r>0$)
when $\lambda \to \infty$. (Persueing this possibility, we will later also conclude that $f$ must tend to the singular limit $f=1$ ($r>0$) based on a series of analytic arguments alone.)
The way this limit is approached may be understood by
plotting $f$ as a function of the scaled radial coordinate $\sqrt{\lambda}
r$. The shape of $f(\sqrt{\lambda}  r)$ depends on $\lambda$ very slightly,
reaching the profile of the limiting case ($\lambda=\infty$) very quickly, above $\lambda
\approx 100$. We show this fact in Fig.~4 where $f$ is plotted
as a function of $\sqrt{\lambda}  r$ for Nielsen-Olesen solutions with $n=1$ and
several values of $\lambda$. The main consequence of this is that the region
 where $f$ differs from 1 for large $\lambda$ has a width
of order $1/\sqrt{\lambda}$.

\begin{figure}[h!]
\epsfxsize=7cm
\centerline{\includegraphics[angle=0,width=92mm]{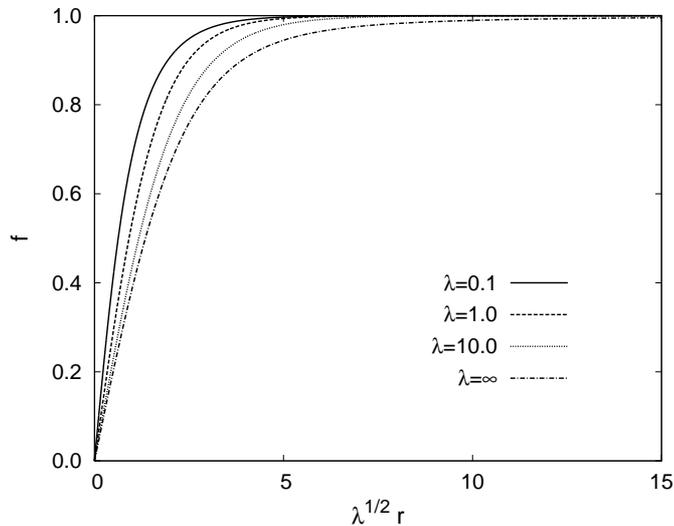}}
\caption{
Function $f$ as a function of $\sqrt{\lambda} r$ for $\lambda=0.1, 1.0, 10.0$
and the limiting case $\lambda=\infty$.
}
\label{fig4}
\end{figure}

In the limit $\lambda \to \infty$, the function $a$ satisfies the equation
\begin{equation}
a'' - \frac{1}{r} a' -a = -n \quad\quad (r>0) \ .
\end{equation}
%in the limit.
The general solution of this equation is
\begin{equation}
a= n + c_1 r K_1 (r) + c_2 r I_1 (r) \ ,
\end{equation}
in terms of the modified Bessel functions $K_1$ and $I_1$. The condition for $r\rightarrow\infty$ implies $c_2 =0$, the condition $a(0)=0$ means $c_1 = -n$, and we have
\begin{equation}
E_3 = \pi n^2 \int_0^\infty r K_1^2 (r) \; dr ,
\end{equation}
which is divergent, since the integrand is of order $1/r$ for small $r$. So
the energy is definitely not bounded in the limit $\lambda \to \infty$.

That $a\rightarrow n-nrK_1 (r)$ and therefore $F_{12}=a'(r)/r=nK_0 (r)$ as $\lambda\rightarrow\infty$ has been shown before by Berger and Chen~\cite{Berger-Chen}. Berger and Chen study the equation for the magnetic field $F_{12}$. They show that the equation for $F_{12}$ linearises and is of the form
\begin{equation}
-\Delta F_{12} ({\vec x}) + F_{12} ({\vec x}) = 2\pi n\delta ({\vec x}) ,
\end{equation}
in the limit $\lambda\rightarrow\infty$. $F_{12}=nK_0 (r)$ is the solution of this equation.

Before we derive the asymptotic behaviour of $E_3$ for large $\lambda$, we calculate the $\lambda\rightarrow\infty$ limit of $E_1$ and $E_4$. For $a=n-nrK_1 (r)$ we have
\begin{equation}
E_1 = \pi n^2 \int_0^\infty r K_0^2 \; dr = \frac{\pi n^2}{2} \left[ r^2 (K_0^2 - K_1^2 ) \right]_0^\infty
= \frac{\pi n^2}{2} .
\end{equation}
Because the solution $(a(r),f(r))$ minimizes the energy, we have a family of
functions $(a(\gamma r),f(\gamma r))$ that satisfies
\[ \left. \frac{d}{d\gamma} E[a(\gamma r),f(\gamma r)] \right|_{\gamma = 1} = \left. \frac{d}{d\gamma}
\left( \gamma^2 E_1 [a(r)] + E_2 [f(r)] + E_3 [a(r),f(r)] + \gamma^{-2} E_4 [f(r)] \right) \right|_{\gamma =1}
\]
\begin{equation}\label{EQ-Derrick}
= 2(E_1 [a(r)] - E_4 [f(r)]) = 0 ,
\end{equation}
which is a manifestation of Derrick's theorem.
Therefore, both $E_1$ and $E_4$ approach the finite value $\pi n^2 /2$ in the
limit $\lambda\rightarrow\infty$, in agreement with the numerical computations
(see Fig.~3).
Using the asymptotic value of $E_4$ in Eq.~(\ref{EQ-dE}), we get
\begin{equation}\label{EQ-EAsym}
\frac{dE}{d\lambda} = \frac{\pi n^2}{2\lambda} \quad\Leftrightarrow\quad
E= \frac{\pi n^2}{2}\log\lambda \ ,
\end{equation}
to leading order. Since the solution minimizes the energy, the second
possibility, where $f$ tends to $f=1$ ($r>0$), must be the one that is realized.
We have already seen that the energy is at least of order $\lambda$ for large $\lambda$,
if $f$ does not tend to $f=1$ ($r>0$).

The logarithmic divergence of the energy of Nielsen-Olesen
solutions in the limit of large $\lambda$ comes from the contribution $E_3$,
since $E_2$ remains finite. $E_2$ and $E_3$ have the following behavior for large $\lambda$:

\begin{eqnarray}
%&&E_1 = E_4 = \frac{\pi}{2} n^2 + o(1) \ , \\
&&E_2 = n^2 \Delta_2(n) + o(1)  \ , \\
&&E_3 = \frac{\pi}{2} n^2 \log \lambda + n^2 \Delta_3 (n) + o(1) \ , \label{behav_Es}
\end{eqnarray}
where the first three values of the functions $\Delta_2(n)$ and $\Delta_3(n)$ are
\begin{equation}
\Delta_2(n) = \left\{
\begin{array}{ll}
0.87679 , & n=1 \\
0.32589 , & n=2 \\
0.17708 , & n=3
\end{array}
\right. \ , \label{Delta2}
\end{equation}
and
\begin{equation}
\Delta_3(n) = \left\{
\begin{array}{ll}
-3.54639 , & n=1 \\
-5.94920 , & n=2 \\
-7.27239 , & n=3
\end{array}
\right. \ , \label{Delta3}
\end{equation}
respectively.

Before we continue with our asymptotic analysis, we look at the variational analysis by Hill et al.~\cite{Hill} for large Ginzburg-Landau parameter. Hill et al. use the functions
\begin{equation}\label{EQ-Hill}
f=1-{\rm e}^{-\mu r} , \quad\quad a=n (1-{\rm e}^{-hr})^2 ,
\end{equation}
and minimize the energy with respect to $\mu$ and $h$. (From our previous discussion we know that $\mu$ should go to infinity and $h$ should go to a constant as $\lambda\rightarrow\infty$, if there is any chance of approximating the correct asymptotic results.) With this ansatz the four terms of the energy are
\begin{equation}
E_1 = 4\pi n^2 h^2 \log\frac{9}{8} , \quad E_2 = \pi /4 , \quad E_3 = \pi n^2 G(s) ,
\quad E_4 = \frac{89 \pi \lambda}{576 \mu^2} ,
\end{equation}
where $s=\mu /h$ and
\begin{equation}
G(s) = \log \frac{3^4 (s+2)^7 (2s+3)^4 (s+4)^2}{2^{11} (s+3)^8 (s+1)^4} .
\end{equation}
Minimizing the energy with respect to $\mu$ and $h$ leads to the equations
\begin{equation}
\frac{h}{\mu^3} = \frac{288 n^2 G'(s)}{89\lambda} , \quad
\frac{h^3}{\mu} = \frac{G'(s)}{8\log (9/8)} .
\end{equation}

For large $s$, $G(s) = \log s + \log (3^4 /2^7 ) + O(1/s^2)$,
\begin{equation}
\mu = \frac{\sqrt{89\lambda}}{12\sqrt{2}n} + O\left(\frac{1}{\sqrt\lambda}\right), \quad\quad
h=\frac{1}{2\sqrt{2\log (9/8)}} + O\left(\frac{1}{\lambda}\right),
\end{equation}
and
\begin{equation}
E_1 = \frac{\pi n^2}{2} , \quad E_2 = \pi /4 , \quad E_3 = \pi n^2
\left( \frac{1}{2} \log\lambda + \log \frac{3^3 \sqrt{89\log (9/8)}}{2^8 n} \right) , \quad E_4 = \frac{\pi n^2}{2},
\end{equation}
up to order o(1). We see that this approximation gives the correct leading
terms for $E_1$ and $E_3$. Using the argument we used in
Eq.~(\ref{EQ-Derrick}) on the energy $E(\mu ,h)$ we get $E_1 = E_4$, and
therefore the leading term of $E_4$ must also be correct. The $O(1)$ terms in
$E_2$, $E_3$ and the total energy $E$ are not correct. For $n=1$, e.g., the
variational method gives the upper bound $E=(\pi /2) \log\lambda + 0.551$,
whereas the correct value is $E=(\pi /2) \log\lambda + 0.472$, as we
saw previously (see Eq.~(\ref{Delta})). That we do not obtain the correct values is no surprise. For
$\lambda\rightarrow\infty$ the function $f$ in Eq.~(\ref{EQ-Hill}) goes to the
step function, which is the correct asymptotic limit. The function $a$ in
Eq.~(\ref{EQ-Hill}), however, does not go to $n-nrK_1 (r)$. Furthermore, the limit
is not approached using the asymptotic expansions of solutions. The functions
in Eq.~(\ref{EQ-Hill}) do not even have the correct asymptotic behaviour
Eq.~(\ref{EQ-Asym2}) for large $r$.

We now give the correct asymptotic approximation for large
$\lambda$. Motivated by Fig. 4 and its interpretation, we are looking for a
family of approximations with the following features: In the outer region, $f$
approaches 1, and $a$ approaches $n-nrK_1$. In the boundary layer (for
$r\lesssim r_0$), $f$ gets
steeper with increasing $\lambda$ and the width of the layer goes to zero in
the limit. This means that the outer approximation $a=n-nrK_1$ extends down to
$r=0$ in the limit $\lambda\rightarrow\infty$, although $n-nrK_1$ does not
have the asymptotic behavior Eq.~(\ref{EQ-Asym1}) of $a$, since $1-rK_1 = -
(r/2)\log r + ... $ for small $r$, i.e., the limit is singular. 

Away from the boundary layer, we look for an outer solution of the form
\begin{equation}\label{f_a_asymp}
f=1 - \frac{1}{\lambda}{\tilde f} + \dots , \quad\quad a = n-nrK_1 + \frac{1}{\lambda}{\tilde a} + \dots
\quad\quad (r>r_0 )
\end{equation}
and find
\begin{equation}
{\tilde f} = n^2 K_1^2 , \quad\quad {\tilde a} = k_n r K_1 + 2n^3 rK_1 \int_r^\infty s I_1 (s) K_1^3 (s) \, ds
-  2n^3 rI_1 \int_r^\infty s K_1^4 (s) \, ds  ,
\end{equation}
where $k_n$ is a constant.
For $r\gg 1$ the solutions are of the form Eq.~(\ref{EQ-Asym2}) with $\alpha =
\sqrt{\pi /2}n$. Also ${\tilde f} /\lambda \ll 1$ holds for $r\gg
1/\sqrt{\lambda}$.

In the boundary layer, $a$ stays very small and $f$ rises rapidly. As an approximation we can therefore use for the inner solution the equation
\begin{equation}
f'' + \frac{1}{r} f' -\frac{n^2}{r^2} f = \frac{\lambda}{2} (f^2 - 1) f \quad\quad (0<r<r_0 )
\end{equation}
with
\begin{equation}
f(0)=0, \quad\quad f(r_0 ) = 1- \frac{1}{\lambda} {\tilde f}(r_0) ,
\end{equation}
instead of using the second-order equation for $f$ in Eq.~(\ref{EQ-af}).
The solution of this boundary value problem, denoted by ${\hat f}$, has to be found numerically. Given ${\hat f}$, we then have to solve the equation
\begin{equation}
a'' - \frac{1}{r} a' + {\hat f}^2 (n-a) = 0
\end{equation}
with
\begin{equation}
a(0)=0, \quad\quad a(r_0 ) = n- nrK_1 (r_0) + \frac{1}{\lambda} {\tilde a}(r_0) .
\end{equation}

\noindent We will denote this inner function $a$ by $\hat a$.

\begin{figure}[h!]
\epsfxsize=7cm
\centerline{\includegraphics[angle=0,width=92mm]{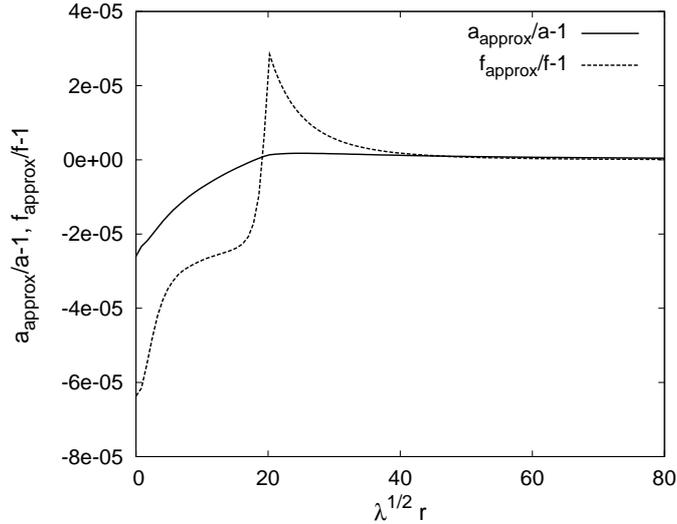}}
\caption{
Comparison of the numerical (exact) functions $a$ and $f$ for $n=1$ and
$\lambda=100$ with the linear
approximations $a_{\rm approx}$ and $f_{\rm approx}$ given by Eq.~(\ref{f_a_asymp}) away from the boundary
layer and $\hat a$ and $\hat f$ in the boundary layer.
}
\label{fig5}
\end{figure}

In order to show that a good linear approximation of the functions $a$ and $f$
for large $\lambda$ is given by Eq.~(\ref{f_a_asymp}) away from the boundary
layer and $\hat a$ and $\hat f$ in the boundary layer, we compare in Fig.~5
the numerical (exact) functions $a$ and $f$ with the corresponding linear
approximations $a_{\rm approx}$ and $f_{\rm approx}$ for $n=1$ and
$\lambda=100$. We observe that for a value of the location of the layer $r_0$
such that $\sqrt{\lambda} r_0 \approx 20$, the relative deviation of the
approximation with respect to the exact values is of the order of
$10^{-5}$. This agreement improves as $\lambda$ is increased.

%%%%%%%%%%%%%%%%%%%%%%%%%%%%%%%%%%%%%%%%%%%%%%%%%%%%%%%%%%%%%%%%%%%%%%%%%%%%%%
\section{Conclusions}
%%%%%%%%%%%%%%%%%%%%%%%%%%%%%%%%%%%%%%%%%%%%%%%%%%%%%%%%%%%%%%%%%%%%%%%%%%%%%%

To complete the study of the four terms which contribute to the energy, we have used numerical computations. The asymptotic result for the total energy Eq.~(\ref{EQ-EAsym}), however, follows from a simple chain of analytic arguments, as we have seen. In contrast to Hill et al.~\cite{Hill} we make no assumptions about the class of functions to be considered. An important step in our chain of arguments is that in the $\lambda\rightarrow\infty$ limit the Higgs field takes its vacuum value for $r>0$. In this regard, the vortex behaves like the monopole~\cite{Kirkman-Zachos}. The crucial difference is that after the Higgs field has decoupled, the energy from the interaction of the Higgs field and the gauge field diverges in the case of vortices, whereas it is finite in the case of monopoles.

%%%%%%%%%%%%%%%%%%%%%%%%%%%%%%%%%%%%%%%%%%%%%%%%%%%%%%%%%%%%%%%%%%%%%%%%%%%%%%

\bigskip
\noindent
{\it Acknowledgement}

\medskip
We are grateful to D. H. Tchrakian for very helpful discussions. F.N.L. gratefully acknowledges Minis\-terio de Ciencia e
  Innovaci\'on of Spain for financial support under Project No. FIS2009-10614. 

%%%%%%%%%%%%%%%%%%%%%%%%%%%%%%%%%%%%%%%%%%%%%%%%%%%%%%%%%%%%%%%%%%%%%%%%%%%%%%


\begin{thebibliography}{99}

\bibitem{Nielsen-Olesen}
  H.~B.~Nielsen and P.~Olesen,
  Nucl.\ Phys.\  B {\bf 61}, 45 (1973).

\bibitem{Tyupkin}
  Y.~S.~Tyupkin, V.~A.~Fateev and A.~S.~Shvarts,
  JETP.\ Lett.\ {\bf 21}, 41 (1975) [Zk.\ Eksp.\ Teor.\ Fiz.\ Pisma\ Red.\ {\bf21}, 91 (1975)].

 
\bibitem{Berger-Chen}
  M.~S.~Berger and Y.~Y.~Chen,
  J.\ Func.\ Anal.\ {\bf 82}, 259 (1989).


\bibitem{Abrikosov}
  A.~A.~Abrikosov,
  Sov.\ Phys.\ JETP.\ {\bf 5}, 1174 (1957) [Zk.\ Eksp.\ Teor.\ Fiz.\ {\bf32}, 1442 (1957)].

\bibitem{Hindmarsh-Kibble}
  M.~B.~Hindmarsh and T.~W.~B.~Kibble,
  Rep.\ Prog.\ Phys.\ {\bf 58}, 477 (1995).

\bibitem{Jacobs-Rebbi}
  L.~Jacobs and C.~Rebbi,
  Phys.\ Rev.\ B {19}, 4486 (1979).

\bibitem{Hill}
  C.~T.~Hill, H.~M.~Hodges and M.~S.~Turner,
  Phys.\ Rev.\ D {\bf 37}, 263 (1988).

\bibitem{Kirkman-Zachos}
  T.~W.~Kirkman and C.~K.~Zachos,
  Phys.\ Rev.\ D {\bf 24}, 999 (1981).

\bibitem{Brihaye 1}
  Y.~Brihaye, C.~T.~Hill and C.~K.~Zachos,
  Phys.\ Rev.\ D {\bf 70}, 111502 (2004).

\bibitem{Brihaye 2}
  Y.~Brihaye, J.~Burzlaff and D.~H.~Tchrakian,
  Phys.\ Rev.\ D {\bf 77}, 107701 (2008).

\bibitem{Perivolaropoulos}
  L.~Perivolaropoulos,
  Phys.\ Rev.\ D {\bf 48}, 5961 (1993).

\bibitem{colsys}
U. Asher, J. Christiansen and R. D. Russel,
Math. Comput. {\bf 33}, 659 (1979);
ACM Trans. Math. Softw. {\bf 7}, 209 (1981).

\bibitem{Taubes}
  C.~Taubes,
  Commun.\ Math.\ Phys.\ {\bf 72}, 277 (1980).


\end{thebibliography}
\end{document}